\documentclass[aps,prl,twocolumn,superscriptaddress,showpacs,floatfix,a4paper]{revtex4}
\usepackage{graphicx}
\usepackage{amsmath}

\begin{document}

\title{Transmission phase of a quantum dot: Testing the role of
population switching}
\author{Moshe Goldstein}
\author{Richard Berkovits}
\affiliation{The Minerva Center, Department of Physics, Bar-Ilan
University, Ramat-Gan 52900, Israel}
\author{Yuval Gefen}
\affiliation{Department of Condensed Matter Physics,
The Weizmann Institute of Science, Rehovot 76100, Israel}
\author{Hans A. Weidenm\"{u}ller}
\affiliation{Max--Planck--Institut f\"{u}r Kernphysik, D-69029 Heidelberg,
Germany}

\begin{abstract}
We propose a controlled experiment to clarify the physical mechanism
causing phase lapses of the amplitude for electron transmission
through nanoscale devices. Such lapses are generically observed in
valleys between adjacent Coulomb--blockade peaks. The experiment
involves two quantum dots embedded in the same arm of an
Aharonov--Bohm interferometer. It offers a decisive test of
``population switching'', one of the leading contenders for an
explanation of the phenomenon.
\end{abstract}


\maketitle

In experiments on the phase $\phi$ of the transmission amplitude
$\mathcal{T}$ through a quantum dot (QD), the following striking pattern
has been observed~\cite{schuster97, avinun05, kobayashi0203}: As a
function of gate voltage $V_g$, $\phi$ increases (as expected) by
$\approx \pi$ over the width of a Coulomb--blockade peak for the
conductance but then (unexpectedly) displays a sharp phase lapse (PL) of
$\approx - \pi$ in the adjacent Coulomb--blockade valley. The PL is found
to occur in every conductance valley between two Coulomb--blockade peaks.
No general consensus as to the mechanism underlying the PL has been
reached yet in spite of determined theoretical efforts, see the
reviews~\cite{hackenbroich01}. Here we propose a controlled
experimental test to confirm or rule out one of the key mechanisms
(``population switching'') considered in the literature.

The need for some mechanism to induce PLs is seen as follows. We
consider transmission through a QD that supports two
orbital~\cite{spin} levels $i = 1,2$. The levels are coupled to
single--channel leads ($\alpha = $ R,L with R,L for right, left,
respectively) by real~\cite{silva02} tunneling matrix elements
$t_{i,\alpha}$. Depending on the value of $s \equiv \prod_{i,\alpha}
t_{i,\alpha} $, we distinguish~\cite{silva02} two cases: $s > 0$ and
$s < 0$. PL is a manifestation of the vanishing of $\mathcal{T}$.
We consider first the case of no electron--electron interaction at zero
temperature. By the Friedel sum rule~\cite{rlsym,datta97}, $\mathcal{T}$
is given by $\exp[i \pi (n_1 + n_2) ]\sin[\pi (n_1 \pm n_2)]$ where
$n_{1,2}$ are the populations of levels $1$ and $2$, respectively, and
the sign is that of $s$. In the valley between two Coulomb--blockade
peaks the lower (upper) level $1$ ($2$) is almost full (empty), and
for $s > 0$ $\mathcal{T}$ vanishes there. For $s < 0$ a PL occurs for
$n_1 = n_2$ and that condition is not met in the valley.  In reality we
expect the signs of the $t_{i \alpha}$ to be random.  Then, correlated
sequences of PLs are not expected, cf. measurements on uncorrelated
mesoscopic QDs~\cite{avinun05}. One faces a similar dilemma
for interacting electrons since the Friedel sum rule is also
valid~\cite{langreth66} in that case. Thus explaining the occurrence
of correlated sequences of PLs implies finding a mechanism by which a
PL occurs for $s < 0$.

\begin{figure}
\includegraphics[width=8cm,height=7cm]{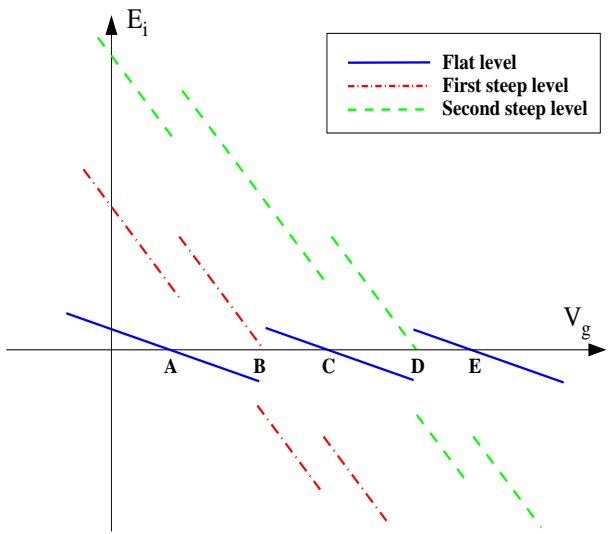}
\caption{\label{fig:scenario}
(Color online) Scenario I (after
Refs.~\onlinecite{hackenbroich97,baltin99a}): The renormalized
(Hartree) energies of the ``flat'' and ``steep'' levels are
schematically shown as functions of the applied gate voltage $V_g$.
In our picture, population switching is discontinuous. As $V_g$
increases, a flat level becomes populated at $V_g = A$. That increases
the energy of the empty steep levels. At $V_g = B$, the lower steep
level crosses the Fermi surface and becomes occupied, causing a
depletion and a rise in energy of the flat level, and population
switching. At $V_g = C$, the flat level is filled again, and the
process repeats itself with the next steep level. We thus obtain a
sequence of population switchings.}
\end{figure}

Population switching provides one such mechanism. It requires the
populations $n_1(V_g)$ and $n_2(V_g)$ of the two levels to become
equal, $n_1(V^{(0)}_g) = n_2(V^{(0)}_g)$, at some value $V^{(0)}_g$ of
$V_g$ in the Coulomb-blockade valley and to switch ($n_2(V_g) >
n_1(V_g)$ for $V_g > V^{(0)}_g$) as $V_g$ is increased
further~\cite{popswitch}. The case $s > 0$ is symmetric in $n_1$ and
$n_2$: PLs appear then irrespective of population switching, while for
$s < 0$ a population switching would produce a PL. Population switching has
been considered in two somewhat different scenarios. The first, displayed
and explained in Fig.~\ref{fig:scenario}, requires two sets of energy
levels which respond differently to $V_g$, a set of ``flat'' and a set
of ``steep'' levels with small (large) slopes, respectively, the
occupancy of which depends non--monotonically on
$V_g$~\cite{hackenbroich97,baltin99a}. In the second scenario (not
displayed) a set of energy levels with identical slopes contains both
broad and narrow levels~\cite{silvestrov00}.  In both scenarios, the
interplay between tunneling and charging gives rise to population
switching. These scenarios have been investigated within a mean--field
approximation~\cite{Gol,goldstein07}, perturbative
calculations~\cite{koenig05}, the numerical renormalization--group
approach for scenario II~\cite{sindel05}, the density--matrix
renormalization--group approach~\cite{berkovits05}, and the functional
renormalization--group (FRG) approach~\cite{karrasch06b}. Either
scenario implies special requirements (e.g., commensurability of the
spacings of the flat set and the steep set in scenario
I or the presence of a generic ultra--broad level in scenario II).
In the sequel we focus attention on the more easily realizable
scenario I.

\begin{figure}
\includegraphics[width=8cm,height=6cm]{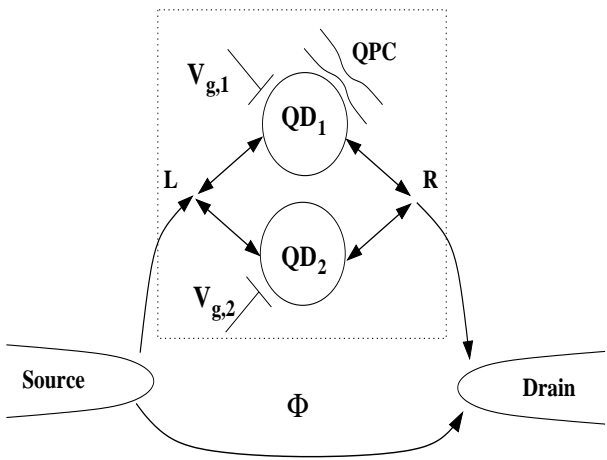}
\caption{\label{fig:doubledot} Schematic view of the proposed setup.
Two quantum dots (QD${_1}$ and QD${_2}$) are embedded into the same
arm of an Aharonov--Bohm interferometer and are connected in parallel
to a left (L) and a right (R) lead. Arrows denote possible tunneling
processes. A quantum point contact (QPC) probes the changes in
population of QD${_1}$. Our analysis addresses the physical processes
within the dotted box.}
\end{figure}

Our proposed experimental setup to test the idea of population
switching is schematically shown in Fig.~\ref{fig:doubledot}. By
varying separately the gate voltages applied to each dot, and by
adjusting the strengths of the dot--lead couplings, one can tune the
levels in one dot independently of those in the other. This makes it
possible to realize both the first and the second scenario mentioned
above, and to tune in and out of the conditions for observing a
correlated sequence of PLs. For scenario I, the two sets of levels are
those in QD${_1}$ and in QD${_2}$, respectively. With the help of our
setup, it is possible to test experimentally the idea that PLs for
$s < 0$ come hand-in-hand with population switching.

To investigate the expected properties of our setup theoretically, we
restrict ourselves to spin--polarized electrons and neglect both
tunneling between the two dots, and the electron--electron interaction
in the leads. The Hamiltonian consists of three parts,
\begin{equation}
{\hat H} = {\hat H}_{D} + {\hat H}_{L} + {\hat H}_{T} \ .
\end{equation}
Here ${\hat H}_D$ is the Hamiltonian for the dots,
\begin{equation} \label{eqn:hd}
{\hat H}_{D} =
\sum_{i=1,2;j} \epsilon_{ij} {\hat a}^{\dagger}_{ij} {\hat a}_{ij} +
\sum_{i=1,2} \frac{U_{i}}{2} \sum_{j \ne j\prime} {\hat n}_{ij}
{\hat n}_{ij\prime} +
U_{12}  \sum_{j; j\prime} {\hat n}_{1j} {\hat n}_{2j\prime} \ ,
\end{equation}
${\hat H}_L$ is the Hamiltonian for the leads,
\begin{equation}
{\hat H}_{L} = \sum_{\ell=L,R;k} \epsilon_{\ell,k}
{\hat c}^{\dagger}_{\ell,k} {\hat c}_{\ell,k},
\end{equation}
and the dot--lead coupling is given by
\begin{equation}
{\hat H}_{T} = \sum_{\substack{i=1,2;j \\ \ell=L,R;k}}
\left( t^{ij}_{\ell,k} {\hat a}^{\dagger}_{ij} {\hat c}_{\ell,k}
+ \mathrm{H.c.} \right)
\end{equation}
while ${\hat a}_{ij}$ (${\hat c}_{\ell,k}$) are the Fermi operators of
the $j$'th level of the $i$'th dot ($k$'th mode of the $\ell$'th lead,
respectively), ${\hat n}_{ij}$ are the number operators, and $\epsilon_{ij} =
\epsilon^{(0)}_{ij}-eV_{g,i}$ are the single--particle energies modified
by the gate voltage. The intra-- and inter--dot charging energies are
denoted by $U_i$ and $U_{12}$, respectively. We assume a constant
density of states in the leads with a band width that exceeds all
other energy scales. The real tunneling matrix elements
$t^{ij}_{\ell,k}$ are taken to be independent of $k$.

In the calculations we use the FRG which has recently been applied to
similar systems~\cite{meden05,karrasch06b}. Earlier calculations using
that method have resulted in accuracy comparable to NRG, at least for
zero temperature and when not more than two levels are close to each
other~\cite{karrasch06b}. These conditions
are met in our case. FRG is based on a functional integral formulation
with an infra--red cutoff. The cutoff dependence of the vertex functions
is given in terms of an exact hierarchy of coupled nonlinear differential
RG equations. For very large values of the cutoff all the modes of the
system are excluded, and the vertex functions are given by the bare
parameters of the Hamiltonian. In principle, the exact vertex
functions could be found by integrating the FRG equations from that
point to the limit where the cutoff tends to zero (in which case all
the modes of the system are included). However, to make the
computation feasible, some truncation scheme must be applied. Usually
one neglects all vertices not present within the bare Hamiltonian,
i.e., three--particle or higher vertex functions, as well as the energy
dependence of the one-- and two--particle vertex functions
\cite{meden05,karrasch06b}. The resulting set of equations can then
be solved numerically. From the (approximate) single--particle vertex
functions the dots' single--particle Green functions, the level
occupations, the linear conductance, and the transmission phase are
readily derived.

For scenario I we assume in the calculation that QD${_1}$ is so small
that only one of its levels plays an active role and functions as the
flat level in Fig.~\ref{fig:scenario}. The levels in QD${_2}$ are
steep and must be well separated to avoid population switching amongst
them, see Refs.~\cite{Gol,silvestrov00,karrasch06b,silvestrov07}.
We vary the gate
voltages on both QDs simultaneously but not at the same pace so as to
induce level crossings. For the gate voltages we write $V_{g,1} =
\alpha V_{g,2} + V_0$ where $0 \le \alpha \le 1$. $V_0$ is chosen so
that the flat level gets filled before it encounters the first steep
level. To estimate $\alpha$ we observe that the change of $V_{g, 2}$
between adjacent crossings of two steep levels with the Fermi surface
is roughly given by $U_2+\Delta_2$, $\Delta_2$ being the mean level
spacing in QD$_2$. As $V_{g, 2}$ is changed, the flat level
must not sink too deeply below the Fermi surface so that it can
eventually get depleted due to the inter--dot interaction of strength
$U_{12}$. That implies that as $V_{g, 2}$ changes by $U_2+\Delta_2$,
$V_{g, 1}$ should change roughly by $U_{1 2}$, so that
$\alpha \approx U_{12}/(U_2+\Delta_2)$.
Too large a value of $\alpha$ will take the flat
level too far down to be depopulated while for too small a value it
will not repopulate. In both these cases, PLs should occur at random,
and the absence of a correlated sequence of PLs should be akin to the
mesoscopic fluctuations of PLs observed in Ref.~\cite{avinun05}, while
for intermediate values of $\alpha$ we expect to see a sequence of
consecutive PLs. Tuning of $\alpha$ to a range which implies population
switching (and consequently the occurrence of PLs) should be
experimentally possible with the aid of the QPC,
Fig.~\ref{fig:doubledot} (The latter is employed to detect the
occurrence of population switching).

\begin{figure}
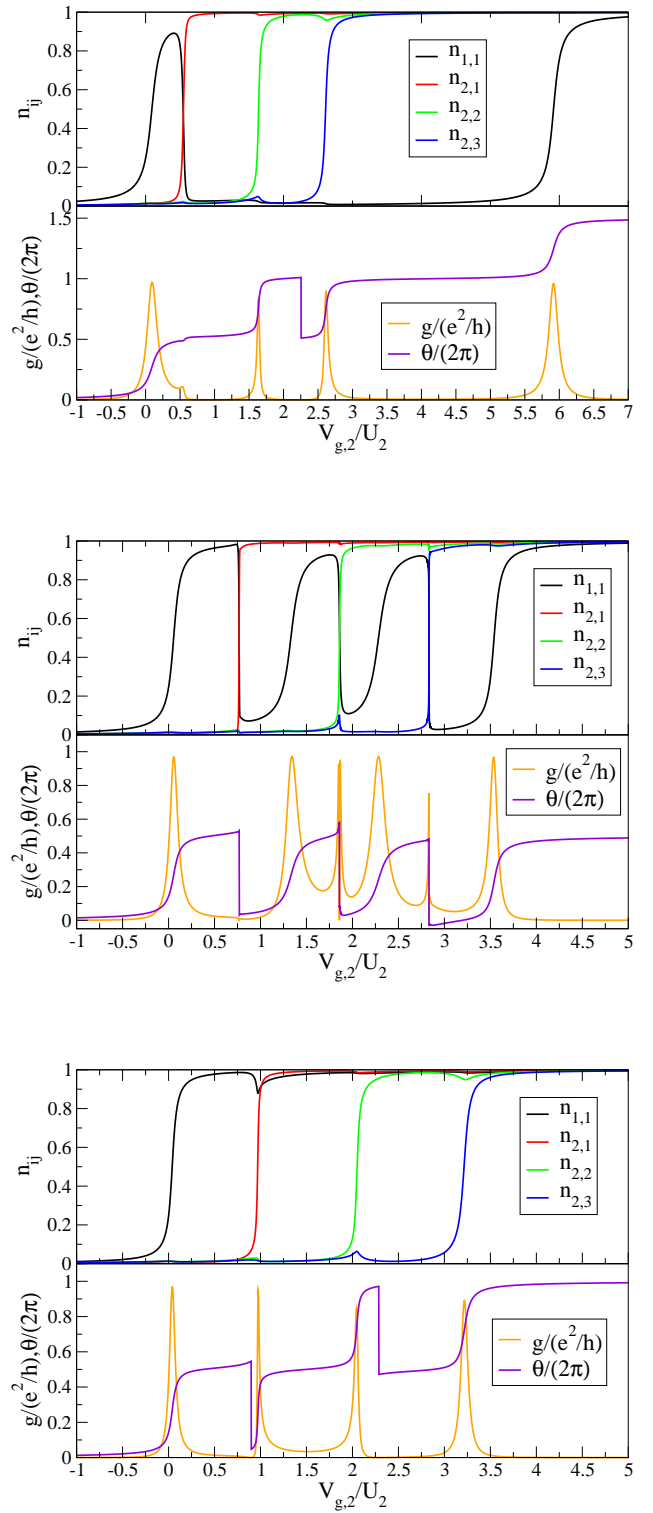

\includegraphics[width=!,height=6cm]{smallalpha}
\vskip 1cm
\includegraphics[width=!,height=6cm]{mediumalpha}
\vskip 1cm
\includegraphics[width=!,height=6cm]{largealpha}
\caption{\label{fig:example} (Color online)
The upper part of each panel shows the population of the flat level
($n_{1,1}$) and of three steep ones ($n_{2,1}$, $n_{2,2}$,
and $n_{2,3}$), the lower part shows
the dimensionless conductance and the transmission phase divided by
$2\pi$, all versus the gate voltage $V_{g,2}$ on QD${_2}$. 
The level energies (in units of $U_2$) are $0$, $0.3$, $0.52$,
and $0.7$, respectively, while their widths due to the coupling to
the left (right) lead are  $0.02$ ($0.03$), $0.018$ ($0.01$),
$0.035$ ($0.016$), $0.039$ ($0.021$).
All the tunneling matrix elements are positive, except
$t^{2,2}_R$ and $t^{2,3}_R$. $U_{12}=0.6U_2$.
The upper, central, and lower panel correspond to
$\alpha=0.3$, $0.5$ [$= U_{12}/(U_2+\Delta_2)$], and $0.7$,
respectively.
}
\end{figure}

\begin{figure}
\includegraphics[width=!,height=6cm]{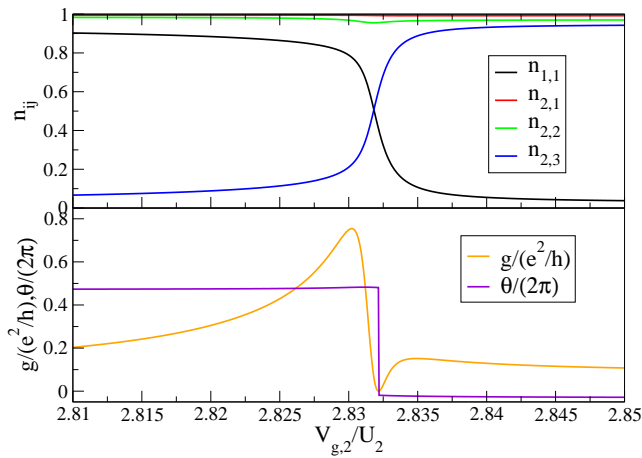}
\caption{\label{fig:zoom} (Color online) Details of one of the PLs
due to population switching ($s < 0$) in the central panel of
Fig.~\ref{fig:example}.}
\end{figure}

This picture is supported by typical results of our calculations shown
in Fig.~\ref{fig:example}. We observe that we obtain a PL in every
Coulomb blockade valley only in the central panel where scenario I
fully applies. Details of the population
switching that occurs in the central panel near $V_{g, 2}/U_2 = 2.83$
in a conductance valley with $s < 0$ are shown in Fig.~\ref{fig:zoom}.
We observe that the population switching is continuous, albeit very
steep. According to Refs.~\cite{silvestrov07}, the scale of the
switching is given by an exponentially small orbital Kondo temperature
$\Delta V_{g,2} \sim T_K = \frac{ \sqrt {U_{12}(\Gamma_1+\Gamma_2)}
}{\pi} \exp \left[ \frac{\pi E_0(U_{12}+\epsilon_0)}{2U_{12}(\Gamma_1
- \Gamma_2)} \ln(\frac{\Gamma_1}{\Gamma_2}) \right]$ \cite{rlsym}. Here
$\Gamma_{1,2}$ are the widths of the two levels that switch population,
and $E_0$ is the average of their positions at the point of population
switching. As expected, a PL occurs in the vicinity of the point of
level crossing.  It is accompanied by two very narrow conductance
peaks. The appearance of these sharp ``correlation--induced
resonances''~\cite{meden05} is easily explained in the case of
left--right symmetry and probably applies at least qualitatively also
for non--symmetric cases. According to the Friedel sum rule, the
conductance is given by $g = (e^2 / h) \sin^2[\pi(n_1-n_2)]$ for $s < 0$,
and is maximal when $|n_1 - n_2| = 1/2$. Since at the population crossing
point $n_1 = n_2$ while far from it either $n_1 = 1$ and $n_2 = 0$ or
$n_1 = 0$ and $n_2 = 1$, conductance peaks should occur on both sides
of the population crossing point. The width of the peaks is again
given by the orbital Kondo temperature. We expect the peaks to
disappear for temperatures higher than that scale~\cite{karrasch06b}.
Similar sharp peaks are seen at the PL near $V_{g, 2}/U_2 = 1.86$
in the central panel of Fig.~\ref{fig:example} but not at the PL
near $V_{g, 2}/U_2 = 0.77$, because there we have $s > 0$.

In summary, we propose an experiment to test the role played by
population switching for phase lapses (PLs) of the transmission
amplitude through a nanoscale device. In a system of two coupled
quantum dots with gate voltages $V_{g, 1}$ and $V_{g, 2}$, we expect
sequences of PLs to occur in consecutive conductance valleys only for
intermediate values of $\alpha = (V_{g,1} - V_0) / V_{g, 2}$. The
associated population switching can be measured by coupling QD${_1}$
to a quantum point contact. Further structures due to
correlation--induced resonances should emerge below the Kondo
temperature and provide an even more detailed test of population
switching. While the present analysis has been focused on scenario I,
very similar phenomena are expected for scenario II.

We thank D.I. Golosov, Y. Oreg and J. von Delft for useful discussions.
M.G. is supported by the Adams Foundation Program of the Israel Academy
of Sciences and Humanities.
Financial support from the Israel Science Foundation (Grants 569/07,
715/08), Schwerpunkt Spintronics SPP 1285, the German-Israeli Foundation
(GIF) and the Minerva Foundation is gratefully acknowledged.


\begin{thebibliography}{99}
\bibitem{spin}
In the present discussion we neglect the spin degrees of freedom. We
do not expect that these will modify our analysis in any qualitative
way \cite{kobayashi0203, stewart97}
\bibitem{rlsym} Here we assume symmetric dot-lead coupling,
$|t_{1L}|=|t_{1R}|, |t_{2L}|=|t_{2R}|$. We expect similar results in
the case of left-right asymmetry.
\bibitem{popswitch} Population switching is central to the two
mechanisms described in Ref.~\cite{Gol}.
\end{thebibliography}
\end{document}